\documentclass[12pt]{article} 
\usepackage[sectionbib]{natbib}
\usepackage{array,epsfig,fancyheadings,rotating}
\usepackage[]{hyperref}  
\usepackage{sectsty, secdot}
\sectionfont{\fontsize{12}{14pt plus.8pt minus .6pt}\selectfont}
\renewcommand{\theequation}{\thesection\arabic{equation}}
\subsectionfont{\fontsize{12}{14pt plus.8pt minus .6pt}\selectfont}

\usepackage{amsmath}
\usepackage{amssymb}
\usepackage{amsfonts}
\usepackage{multirow}
\usepackage{amsthm}

\usepackage{subcaption}
\usepackage{bm}
\usepackage{graphicx}
\usepackage{booktabs}
\usepackage[flushleft]{threeparttable}

\newtheorem{theorem}{Theorem}

\newtheorem{proposition}{Proposition}
\theoremstyle{definition}

\newtheorem{assumption}{Assumption}

\newtheorem{remark}{Remark}
\def\T{{ \mathrm{\scriptscriptstyle T} }}
\def\Pstar{\mathrm{pr}^*}
\def\Estar{\mathrm{E}^*}

\def\op{o_{\mathrm{pr}^*}}
\def\Op{O_{\mathrm{pr}^*}}
\def\setS{\mathcal{S}}

\def\bbeta{\bm{\beta}}
\def\bx{\bm{x}}
\def\sumn{\sum_{i=1}^n}

\DeclareMathOperator*{\argmin}{\arg\,\min}

\newenvironment{asptref}[1]
  {\edef\theassumption{\getrefnumber{#1}\unexpanded{$'$}}\assumption}
  {\endassumption}

\begin{document}


\renewcommand{\baselinestretch}{2}

\markright{ \hbox{\footnotesize\rm 
}\hfill\\[-13pt]
\hbox{\footnotesize\rm
}\hfill }

\markboth{\hfill{\footnotesize\rm Yuanzhi Li and Xuming He} \hfill}
{\hfill {\footnotesize\rm Posterior inference for quantile regression} \hfill}

\renewcommand{\thefootnote}{}
$\ $\par


\fontsize{12}{14pt plus.8pt minus .6pt}\selectfont \vspace{0.8pc}
\centerline{\large\bf Posterior Inference for Quantile Regression:}
\vspace{2pt} 
\centerline{\large\bf Adaptation to Sparsity}
\vspace{.4cm} 
\centerline{Yuanzhi Li and Xuming He} 
\vspace{.4cm} 
\centerline{\it Department of Statistics, University of Michigan}
 \vspace{.55cm} \fontsize{9}{11.5pt plus.8pt minus.6pt}\selectfont


\begin{quotation}
\noindent {\it Abstract:}
Quantile regression is a powerful data analysis tool that accommodates heterogeneous covariate-response relationships.
We find that by coupling the asymmetric Laplace working likelihood with appropriate shrinkage priors, we can deliver posterior inference that automatically adapts to possible sparsity in quantile regression analysis. After a suitable adjustment on the posterior variance, the posterior inference provides asymptotically valid inference under heterogeneity. Furthermore, the proposed approach leads to oracle asymptotic efficiency for the active (nonzero) quantile regression coefficients and super-efficiency for the non-active ones.
By avoiding the need to pursue dichotomous variable selection, the Bayesian computational framework demonstrates desirable inference stability with respect to tuning parameter selection. 
Our work helps to uncloak the value of Bayesian computational methods in frequentist inference for quantile regression. 

\vspace{9pt}
Shrinkage prior,
Working likelihood.
\par
\end{quotation}\par

\def\thefigure{\arabic{figure}}
\def\thetable{\arabic{table}}

\renewcommand{\theequation}{\thesection.\arabic{equation}}

\fontsize{12}{14pt plus.8pt minus .6pt}\selectfont

\section{Introduction}
Quantile regression, formally introduced by \citet{koenker1978regression}, has become a powerful tool for data analysis in a wide range of applications, from economics \citep{fitzenberger2013economic} to public health \citep{wei2019applications}. Quantile regression enables researchers to go beyond the modeling of conditional means; by modeling the effect of covariates at different conditional quantile levels of a response variable, we obtain more comprehensive information on the relationships between the response and the covariates. In particular, quantile regression enables us to reveal the differential effects of a covariate on the low and high end of the response distribution. 

Because the sampling distributions of the quantile regression estimators involve the conditional density functions as nonparametric nuisance parameters, inferential methods have to approximate those quantities directly or indirectly. Existing methods include the use of plugged-in density estimates \citep{powell1991estimation,hendricks1992hierarchical}, rank-score tests \citep{gutenbrunner1993tests,koenker1999goodness}, re-sampling methods \citep{feng2011wild,pan2020multiplier}, and Bayesian computational approaches \citep{chernozhukov2003mcmc, yang2016posterior}.

The present paper follows the Bayesian computational framework to deliver frequentist inference for quantile regression. We investigate the asymptotic properties of posterior inference in a possibly sparse model and demonstrates desirable efficiency and stability of the approach.
We show that the pseudo-Bayesian approach based on a working likelihood and a shrinkage prior achieves automatic adaptation to sparsity, and it can provide asymptotically valid inference for quantile regression under heterogeneity. In the rest of our paper, we shall refer to our proposed pseudo-Bayesian approach as a ``Bayesian approach'', even though we pursue inference in the frequentist sense.

More specifically, we consider the asymmetric Laplace working likelihood  \citep{yu2001bayesian, yang2016posterior} with appropriate shrinkage priors in the spirit of common frequentist penalty functions \citep{wu2009variable}. With a random sample of size $n$ from a linear quantile regression model with $p \leq n$ covariates but only $s \le p$ active (non-zero) coefficients, our work offers the following insights into the posterior inference.
 
\vspace{3mm}
 
 1. The posterior distribution concentrates around the true quantile regression parameters at an adaptive rate: it achieves the $n^{-1/2}$ rate for the active coefficients and a super-efficient rate of $o(n^{-1/2})$ for the inactive (zero-valued) coefficients.
 
 2. The posterior mean for the active coefficients is asymptotically normal and oracle efficient: it achieves the same asymptotic variance as the quantile regression estimator as if we knew which coefficients are active/inactive.
 
 3. With an appropriate adjustment of the posterior variance, we can construct automatically adaptive confidence intervals in the frequentist sense: they are asymptotically oracle for the active coefficients, while they are super-efficient for the inactive coefficients with coverage probabilities tending to one.
 
 4. Even if we identify the active covariates correctly, optimally weighted quantile regression estimators cannot be obtained by focusing on only those active covariates. The Bayesian approach does not rely on any binary selection of active/inactive covariates; thus, it tends to offer performance advantages over variable selection approaches. 
 
\vspace{3mm}

It is important to note that unadjusted Bayesian inference is not automatically valid since the posterior is constructed operationally from a mis-specified asymmetric Laplace working likelihood.
Even for finite-dimensional models without the use of shrinkage priors, the posterior distribution does not approximate the sampling distribution of the classical quantile regression estimator \citep{sriram2015sandwich,yang2016posterior}.
However, from the frequentist perspective, we find a relatively simple adjustment to the posterior variance that facilitates asymptotically valid and efficient inference in possibly sparse quantile regression models. 
The Bayesian computational framework allows us to circumvent the nonparametric estimation of the conditional density functions as nuisance parameters \citep{chernozhukov2003mcmc}; therefore, it serves as a valuable tool to frequentist inference.


Bayesian modeling with shrinkage priors has been quite well studied in terms of estimation accuracy (error rates) of the parameters and variable selection in high-dimensional problems; see, e.g., \citet{song2017nearly,jiang2019bayesian} and \citet{gao2020general}. 
The focus of the present paper is not estimation or variable selection consistency but the understanding of what can be accomplished in inference for possibly sparse quantile regression models, about which relatively little has been available in the literature even when the number of predictors $p$ is fixed.
The main challenge is adjusting for the mis-specification of the likelihood function under heterogeneity and model sparsity.
Our work also gives the first asymptotic analysis, as far as we know, for the posterior mean and variance in the Bayesian quantile regression framework with a shrinkage prior. 
To simplify the technicalities and focus on the main points, we begin by working with the asymptotic framework where the sample size $n$ goes to infinity yet the covariate-dimension $p$ is kept fixed; We discuss an extension to the regime where $p$ grows with $n$ at a controlled rate later in the paper.

The key to understanding Bayesian approaches is the combination of likelihood and prior. Since the quantile regression model does not assume any parametric likelihood function, it is common to rely on a working likelihood to pursue posterior inference. Examples of other working likelihoods include the empirical likelihood \citep{yang2012bayesian,xi2016bayesian}, the score likelihood \citep{wu2021bayesian}, or the approximate likelihood \citep{feng2015bayesian}. 
The use of different shrinkage priors is also prevalent in practice for more efficient estimation \citep{li2010bayesian,adlouni2018regularized,kohns2020horseshoe}. This paper adopts the asymmetric Laplace working likelihood and focuses on two easy-to-understand examples of shrinkage priors for their interpretability and computational attractiveness. Under this Bayesian framework, we provide new theoretical insight on the adaptation of posterior inference in possibly sparse quantile regression models. In particular, we propose new adaptive adjustment to the posterior variance that generalizes \citet{yang2016posterior} and provide thorough theoretical investigation of its properties.

The rest of the paper is organized as follows. In Section \ref{sec::setup}, we discuss the quantile regression problem and our Bayesian framework.
Then we give the corrected posterior inference approach in Section \ref{sec::main}, supported by the asymptotic properties of the posterior distribution.
In Section \ref{sec::increase-p}, we discuss the theoretical extension towards the asymptotic regime with moderately increasing covariate-dimension. 
Section \ref{sec::simulation} shows some simulation results to demonstrate the effectiveness and stability of the proposed approach. We make some concluding remarks in Section \ref{sec::conclude}.

\section{Problem setup}
\label{sec::setup}
\subsection{The quantile regression model}

Let $Q_\tau(Y\mid X =\bm{x})$ be the $\tau$th conditional quantile of the response variable $Y$ given covariates $X = \bm{x}$, where $\bm{x}= (x_0,\ldots,x_p)^{\T}$ includes an intercept term $x_0 = 1$ and $p$ covariates, and  $\tau\in(0,1)$ is a pre-specified quantile level of interest. We consider the linear quantile regression model
\begin{equation}
\label{eq::truemodel}
Q_\tau(Y\mid X = \bm{x}) = \bm{x}^{\T}\bbeta^0(\tau),
\end{equation}
where $\bbeta^0(\tau) = (\beta_0^0(\tau),\ldots,\beta_p^0(\tau))^{\T}$ is the true quantile regression coefficient vector.
The conditional median of $\tau =0.5$ is a special case, and high or low quantile levels of $\tau$ are often of interest in applications, ranging from financial risk quantification \citep{taylor2019forecasting} to public health assessment \citep{wei2019applications}. Since we focus on a fixed $\tau$ in the model, we often suppress the index $\tau$ in $\bbeta^0(\tau)$ hereafter.

In this paper, we consider Model (\ref{eq::truemodel}) to be possibly sparse.
Let $\mathcal{S} = \{0\}\cup \{ j \in \{1,\ldots p\}:\; \beta_j^0 \neq 0 \}$ be the index set of the active (non-zero) coefficients, including the intercept term; whereas $\mathcal{S}^c = \{0,\ldots,p\}\backslash \mathcal{S}$ is for the inactive coefficients. Let $s = |\mathcal{S}| - 1$ be the number of active covariates. A possibly sparse model refers to $0 \leq s \leq p$ for some integer $s$; yet we do not know $\mathcal{S}$ in advance.
For now, we suppose the covariate-dimension $p$ is fixed, and discuss an extension towards the case when $p$ can increase in Section \ref{sec::increase-p}.

Here we briefly review the classical quantile regression analysis.
Let $\mathbb{D}_n = \{(\bx_i,y_i):i=1,\ldots,n\}$ be a random sample of size $n$ that satisfies Model (\ref{eq::truemodel}). 
The quantile regression estimator \citep{koenker1978regression} is
\begin{equation}
\label{eq::quantobj}
\widehat{\bbeta} = \argmin_{\bm{u}\in\mathbb{R}^{(p+1)}}\;\sum_{i=1}^n \rho_\tau(y_i - \bx_i^{\T}\bm{u}),
\end{equation}
where $\rho_\tau(v) = v\{\tau - 1(v<0)\}$ and $1(\cdot)$ is the indicator function.
With $p \ll n$, statistical inference for Model (\ref{eq::truemodel}) can be carried out based on the asymptotic properties of the estimator $\widehat{\bbeta}$; we refer to \citet{koenker_2005} and \citet{koenker2017handbook} for more details on quantile regression. Here we highlight two aspects for the estimator $\widehat{\bbeta}$: (i) it does not account for the possible model sparsity, therefore it does not achieve the optimal efficiency when Model (\ref{eq::truemodel}) is sparse; (ii) its asymptotic variance-covariance matrix involves the conditional density function of $Y$ given $X$, which requires non-parametric estimation and can be unstable in practice.



\subsection{A Bayesian framework}
\label{sec::bayes-model}
In this Section, we give the Bayesian framework for modeling the quantile regression coefficient $\bbeta$ in Model (\ref{eq::truemodel}).
We adopt the asymmetric Laplace working likelihood:
\begin{equation}
\label{eq::ALlikelihood}
\mathcal{L}(\mathbb{D}_n \mathrel{\mid} \bbeta) \propto 
\exp\left\{ -\sum_{i=1}^n\rho_\tau(y_i - \bx_i^{\T}\bbeta) \right\},
\end{equation}
where $\propto$ means equality up to a multiplicative factor that does not depend on $\bbeta$. 
We call $\mathcal{L}(\mathbb{D}_n | \bbeta)$ a working likelihood because it does not correspond to the true data-generating mechanism of $\mathbb{D}_n$ under parameter value $\bbeta$; in fact, there is no `true' likelihood function as Model (\ref{eq::truemodel}) itself does not fully specify a conditional distribution of $Y$ given $X$. 
Choosing a working likelihood in the form of (\ref{eq::ALlikelihood}) enjoys two benefits: (i) it allows the maximum working likelihood estimator to coincide with the classical quantile regression estimator $\widehat{\bbeta}$ in (\ref{eq::quantobj}); (ii) 
its Fisher information matrix shares a critical component with the variance-covariance matrix of $\widehat{\bbeta}$ \citep{yang2016posterior}.

\begin{figure}[hbt]
\centering
\includegraphics[width = .85\textwidth]{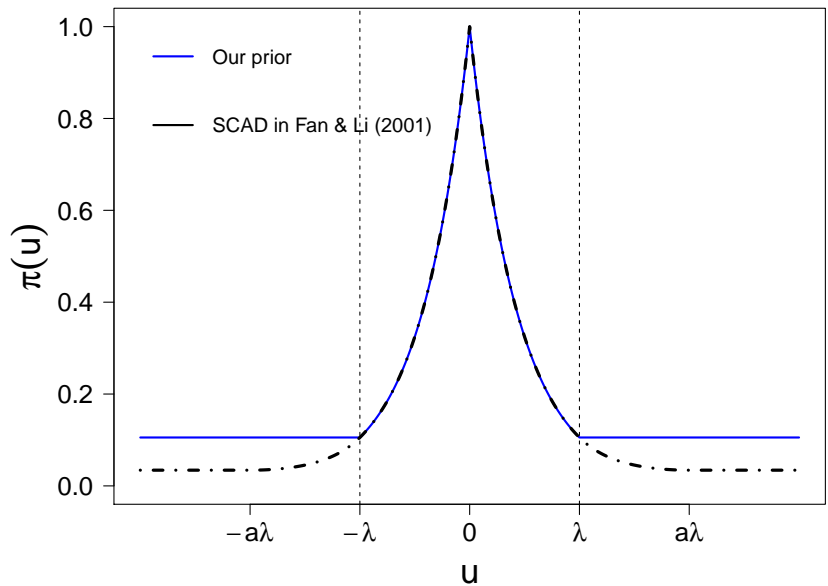}
\caption{Comparison between the prior $\pi_{CA}(u)$ and the prior induced by the SCAD penalty in \citet{fan2001variable}; $a$ is a tuning parameter in the SCAD penalty and we set $a = 2$ in the plot. Both priors are flat when $|u| > a\lambda$.
}
\label{fig::SCADprior}
\end{figure}

To incorporate the possible model sparsity, in this paper we consider two examples of shrinkage priors in the spirit of common penalty functions:
\begin{eqnarray}
\pi_{AL}(\bbeta) &\propto& \exp\left\{ - n^{1/2}\lambda_n\sum_{j=1}^p w_j\lvert \beta_j \rvert \right\}, \label{eq::priorAlasso}\\
\pi_{CA}(\bm{\beta}) &\propto& \exp\left\{ - n\sum_{j=1}^p  p_{\lambda_n}(\beta_j) \right\},\label{eq::priorSCAD}
\end{eqnarray}
where $w_j$ and the function $p_{\lambda_n}(\cdot)$ will be given below; the tuning parameter $\lambda_n$ depends on the sample size, but the subscript $n$ is often suppressed in the paper when there is no confusion. 
The prior (\ref{eq::priorAlasso}) corresponds to the Adaptive Lasso (AL) penalty \citep{zou2006adaptive}, where $w_j = 1/\lvert \hat{\beta}_j\rvert$ for $j\in\{1,\ldots,p\}$ as in \citet{wu2009variable} and $\hat{\beta}_j$ is the $j$-th component of $\widehat{\bbeta}$ defined in (\ref{eq::quantobj}). 
The Clipped Abosolute (CA) prior (\ref{eq::priorSCAD}) is motivated from the Smoothly Clipped Absolute Deviation (SCAD) penalty of \citet{fan2001variable}, where we define the penalty function $p_{\lambda}(u) = \lambda(|u|\wedge \lambda)$. Our choice of $p_{\lambda}(u)$ in (\ref{eq::priorSCAD}) is in the same spirit of the SCAD penalty, but we remove the smoothing component to simplify the theoretical derivation; See Figure \ref{fig::SCADprior} for a visual comparison.
Note for either (\ref{eq::priorAlasso}) or (\ref{eq::priorSCAD}), we impose an improper flat prior for the intercept $\beta_0$, i.e, $\pi(\beta_0) \propto 1$; therefore $\beta_0$ is not penalized.

Given the working likelihood (\ref{eq::ALlikelihood}) and any prior $\pi(\bbeta)$, we have the formal posterior density:
\begin{equation}
\label{eq::posterior_1}
p\left(\bbeta \mathrel{\mid} \mathbb{D}_n\right) \propto \mathcal{L}(\mathbb{D}_n \mathrel{\mid} \bbeta)\times \pi(\bbeta).
\end{equation}
Under either the AL (\ref{eq::priorAlasso}) or CA (\ref{eq::priorSCAD}) prior, existing Markov Chain Monte Carlo (MCMC) algorithms enable efficient sampling from the posterior; See \citet{li2010bayesian} and \citet{alhamzawi2012bayesian} for the prior (\ref{eq::priorAlasso}); \citet{li2011bayesianSCAD} and \citet{adlouni2018regularized} for priors similar to (\ref{eq::priorSCAD}). In the rest of the paper, we shall examine the asymptotic properties of the posterior distribution, from which we derive valid and adaptive confidence intervals in the frequentist sense.


\begin{remark}
Priors (\ref{eq::priorAlasso}) and (\ref{eq::priorSCAD}) are both examples from a wide range of shrinkage priors, which aims to shrink smaller coefficients towards zero while keep those larger coefficients unbiased; See e.g., \citet{song2017nearly} and \citet{jiang2019bayesian} for more discussions in the linear regression context.
In this paper we shall only focus only on two examples to simplify theory and computation for quantile regression, through which we hope to shed some light on the use of shrinakge priors for efficient statistical inference in quantile regression. 
\end{remark}

\section{Adaptive posterior inference}

\label{sec::main}

While posterior inference seems straightforward with a Bayesian framework, its validity is not warranted since the working likelihood is mis-specified \citep{yang2016posterior}.
In this section, we begin by investigating the asymptotic properties of the posterior distribution from the frequentist perspective.
Next, we propose an adjustment to the posterior variance-covariance matrix and show that it can lead to valid confidence intervals that also adapts to model sparsity.
In the last subsection, we discuss an extension of using a weighted working likelihood to obtain the optimal efficiency.
Throughout this section, the covariate-dimension $p$ is fixed in Model (\ref{eq::truemodel}).

\subsection{Notation}
\label{sec::notations}
Recall that we have $\bbeta^0 = (\beta_0^0,\ldots,\beta_p^0)^\T$ as the true regression coefficient in Model (\ref{eq::truemodel}), and that $\mathcal{S}$ is the index set of the active (non-zero) coefficients, including the intercept term.
Without loss of generality, we assume $\setS = \{0, 1,\ldots,s\}$. 
Recall $\widehat{\bbeta}$ is the classical quantile regression estimator in (\ref{eq::quantobj}); let $\tilde{\bbeta}_{\setS}\in\mathbb{R}^{s+1}$ be the oracle quantile regression estimator, which solves (\ref{eq::quantobj}) using only the active covariates. 
For any vector $\bm{v} = (v_0,\ldots,v_p)^\T$, let $\bm{v}_\mathcal{S}= \{v_j:\;j\in\mathcal{S}\}$ and $\bm{v}_{\mathcal{S}^c} = \{v_j:\;j\not\in\mathcal{S}\}$. For any matrix $A \in\mathbb{R}^{(p+1)\times (p+1)}$, we partition 
\[
A = 
\left(
\begin{array}{cc}
A_{\setS}&A_{\setS,\setS^c}\\
A_{\setS^c,\setS}&A_{\setS^c}
\end{array}
\right),
\]
where $A_{\setS} \in \mathbb{R}^{(s+1)\times (s+1)}$; for $i,j\in\{0,\ldots,p\}$, we shall write $A(i,j)$ as the $(i+1,j+1)$th entry of A.

Recall that $\mathbb{D}_n$ contains a random sample of size $n$ from the distribution $(X,Y)\sim\Pstar$ whose $\tau$th conditional quantile of $Y$ satisfies Model (\ref{eq::truemodel}). We will also use $\Estar(\cdot)$ as the expectation operator under $\Pstar$. Let $\epsilon = Y - X^{\T}\bbeta^0$, and  $f_{\epsilon\mid X}$ (or $f_{\epsilon\mid X_{\setS}}$) be the conditional density function of $\epsilon$ given $X$ (or $X_{\setS}$).
Furthermore, let $D = \Estar(XX^{\T})$ and $G = \Estar\{XX^{\T}f_{\epsilon\mid X_{\setS}}(0)\}$.
Given the data $\mathbb{D}_n$ and the prior $\pi$, we consider the posterior probability measure as
\[
 \Pi\left(A\mathrel{\mid} \mathbb{D}_n\right)  = \int_A p(\bbeta \mathrel{\mid} \mathbb{D}_n)\,\mathrm{d}\bbeta,
\] 
for any measurable set $A\subset\mathbb{R}^{(p+1)}$, where  $p(\bbeta \mid \mathbb{D}_n)$ is the posterior density in (\ref{eq::posterior_1}).

We also use the following set of notations in the paper. 
For a vector $\bm{v}$, let $\lVert \bm{v} \rVert$ and $\lVert \bm{v}\rVert_\infty$  be its $\ell_2$ norm and its maximum norm, respectively. 
For a matrix $A$, we denote its maximal/minimal eigenvalue by $\theta_{\max}(A)$ and $\theta_{\min}(A)$, respectively.
For probability density functions $h(x)$ and $g(x)$, we denote their total variation distance by $\lVert h-g \rVert_{TV}=\int |h-g|\mathrm{d}x$. For covariance matrices $A$ and $B$, we write $A\preceq B$ if $B-A$ is positive semi-definite.
For two deterministic sequences $a_n$ and $b_n$, we write $a_n \ll b_n$ if $a_n=o(b_n)$ and $a_n \lesssim b_n$ if there exists a universal constant $C_1 > 0$ such that $a_n \leq C_1b_n$.
For stochastic sequences $A_n$ and $B_n$, we use the notations $A_n \ll_{\Pstar} B_n$ and $A_n \lesssim_{\Pstar}B_n$ to denote $A_n = \op(B_n)$ and $A_n  = \Op(B_n)$, respectively; we define $A_n \asymp_{\Pstar} B_n$ if both $A_n = \Op(B_n)$ and $B_n = \Op(A_n)$ hold.


\subsection{Posterior asymptotics}
\label{subsec::theory}
In this subsection, we present the large-sample properties of the posterior distribution defined in (\ref{eq::posterior_1}).  To this end, we need the following technical assumptions.

\begin{assumption}[Identification]
\label{aspt::id}
For any $\delta > 0 $, there exists $\varepsilon > 0$, such that
\[
\lim\limits_{n\to \infty}\;\Pstar\left[
\sup_{\bbeta:\lVert \bbeta - \bbeta^0\rVert \geq \delta} \left\{\frac{L_n(\bbeta^0) - L_n(\bbeta)}{n}\right\} \leq -\varepsilon
\right]
= 1,
\]
where $L_n(\bbeta) = \sumn \rho_\tau(y_i - \bx_i{^\T}\bbeta)$. 
\end{assumption}

\begin{assumption}[Covariates]
\label{aspt::covar}
The covariate-vector $X$ has bounded support on $\mathcal{X}\subset \mathbb{R}^{p+1}$. Furthermore, the eigenvalues of $D = \Estar(XX{^\T})$ are all bounded away from $0$ and $+\infty$.
\end{assumption}

\begin{assumption}[Conditional densities]
\label{aspt::density}
The conditional density function of $\epsilon = Y - X^\T\bbeta^0$ given $X=\bx$ satisfies: (i)  
there exists $L > 0$ such that for all $u,u'\in\mathbb{R}$,
\[
\sup_{\bx \in\mathcal{X} }|f_{\epsilon\mid X = \bx}(u) - f_{\epsilon\mid X = \bx}(u')| \leq L|u - u'|;
\]
and (ii) there exist two constants $\underline{f} $ and $\overline{f}$, such that 
\[
0< \underline{f} \leq \inf_{\bx \in\mathcal{X}} \left\{f_{\epsilon\mid X=\bx}(0) \right\} \leq 
\sup_{\substack{u\in\mathbb{R}\\\bx\in\mathcal{X}}} \left\{ f_{\epsilon\mid X=\bx}(u)\right\} \leq \overline{f}.
\]
\end{assumption}


\begin{assumption}[Separation]
\label{aspt::sparsity}
For some constant $b_0 >0 $, we have
\begin{gather*}
\min_{j\in\setS\setminus \{0\}}\;\lvert \beta_{j}^0\rvert  > b_0.
\end{gather*}
\end{assumption}

We briefly discuss the assumptions. Assumptions \ref{aspt::id}--\ref{aspt::density} are standard in Bayesian modeling with a working likelihood \citep{chernozhukov2003mcmc,yang2016posterior} and the quantile regression literature \citep{knight1998limiting,pan2020multiplier}; see also \citet[Section 4]{koenker_2005}. 
In particular, the two assertions in Assumption \ref{aspt::density} hold for the conditional density $f_{\epsilon\mid X_{\setS}}(u)$ as well; Furthermore, Assumption \ref{aspt::covar} implies the eigenvalues of $G = \Estar\{XX^{\T}f_{\epsilon\mid X_{\setS}}(0)\}$ are also bounded. Assumption \ref{aspt::sparsity}  holds automatically when we posit a fixed model as (\ref{eq::truemodel}) where $p$ is a constant;
Similar separation conditions are needed to achieve consistent model selection \citep{fan2001variable,wu2009variable,belloni2011}. 

Now we present the main theoretical result regarding the posterior distribution defined in (\ref{eq::posterior_1}).

\begin{theorem}
\label{thm::Unified}
Consider the posterior distribution under either the AL prior (\ref{eq::priorAlasso}) or the CA prior (\ref{eq::priorSCAD}). Suppose Assumptions \ref{aspt::id}--\ref{aspt::sparsity} hold, and the tuning parameter $\lambda$ satisfies $n^{-1/2} \ll \lambda\ll 1$, then we have the following results.
\begin{enumerate}
\item Adaptive rate-of-contraction: for any sequence $M_n\to+\infty$,
\[
\Pi\left( \left\lVert\bbeta_{\setS} - \bbeta^0_{\setS}\right\rVert \leq \frac{M_n}{n^{1/2}} \,,\, \left\lVert\bbeta_{\setS^c} \right\rVert_{\infty} \leq \frac{M_n}{n\lambda} ~\middle\vert~ \mathbb{D}_n\right) \to 1,
\]
in $\Pstar$-probability.
\item Distributional approximation: for some density functions $\pi_j(u) = \Op(1)$ $(u\in\mathbb{R}, j\in\setS^c)$,
\[
\left\lVert p\left(\bbeta ~\middle\vert~  \mathbb{D}_n\right) -  \phi\left(\bbeta_S;\tilde{\bbeta}_S,\frac{1}{n}G_{\setS}^{-1}\right)\times \prod_{j\not\in\setS} \left\{ n\lambda  \pi_j(n\lambda \beta_j)\right\}\right\rVert_{TV} \to 0,
\]
in $\Pstar$-probability,
where $\phi(\cdot\,;\,\mu,\Sigma)$ is the density function of a multivariate-Gaussian distribution. In particular, $\pi_j(u) = (n^{-1/2}w_j/2)\exp\{-n^{-1/2}w_j|u|\}$ if we use the AL prior (\ref{eq::priorAlasso}), and  $\pi_j(u) = (1/2)\exp\{-|u|\}$  if we use the CA prior (\ref{eq::priorSCAD}).
\end{enumerate}
\end{theorem}

Theorem \ref{thm::Unified} shows that, despite the likelihood mis-specification, the posterior under either prior can separate the active and inactive coefficients with a wide range of choices of $\lambda$.
With $n\lambda \gg n^{1/2}$, part 1 of Theorem \ref{thm::Unified} shows the posterior for the inactive coefficients concentrates towards $0$ at a second-order rate, which is super-efficient.
Furthermore, part 2 of Theorem \ref{thm::Unified} shows the posterior for $\bbeta_{\setS}$ and $\bbeta_{\setS^c}$ are approximately independent. In particular, the posterior for $\bbeta_{\setS}$ is  `oracle', i.e., the Gaussian limiting posterior for $\bbeta_{\setS}$ is the same as if we knew the true model $X_\setS$ in advance \citep{sriram2015sandwich} regardless of the prior we use.
Thus, with the two shrinkage priors in Section \ref{sec::bayes-model}, the posterior distribution can automatically adapt to the model sparsity.

Although slightly different in the limit, the posterior shares the same adaptation principle under both the AL and CA priors in Section \ref{sec::bayes-model}.
For an active coefficient, asymptotically the prior casts no effect on the posterior distribution; For an inactive coefficient $\beta_j (j\in\setS^c)$, the shrinkage prior dominates over the working likelihood since the limiting posterior density $n\lambda\times\pi_j(n\lambda\beta_j)$ is proportional to the corresponding prior for $|n\lambda \beta_j| = O(1)$. Therefore, the shrinkage prior can separate the inactive coefficient from those active ones. This phenomena is in line with that in the Gaussian linear model setting under general shrinkage priors \citep[Theorem 2.4]{song2017nearly}.

\subsection{Confidence intervals from posterior moments}
\label{subsec::inference}
Since the working likelihood (\ref{eq::ALlikelihood}) is likely mis-specified, the posterior needs to be properly calibrated to deliver valid frequentist inference for quantile regression. However, the correction on the posterior variance proposed in \citet{yang2016posterior} is no longer valid with the use of shrinkage priors. In light of Theorem \ref{thm::Unified}, we give a modified adjustment that yields confidence intervals based on posterior moments that are automatically adaptive to model sparsity.

We construct the confidence intervals for $\bbeta^0$ based on the posterior mean
$\check{\bbeta} = (\check{\beta}_0,\ldots,\check{\beta}_p), $
and posterior variance-covariance matrix $\check{\Sigma}$  obtained from any posterior sampling algorithm.  We start from the adjustment used in \citet{yang2016posterior} by letting
$\widehat{D} = \sum_{i=1}^n\bx_i\bx_i^{\T}/n$,
 and   $\check{\Sigma}_{adj} = {n\tau(1-\tau)}\check{\Sigma}\,\widehat{D}\,\check{\Sigma}$.
Then our proposed level $1-\alpha$  confidence interval for each $\beta_j^0$ takes the form
\begin{equation}
\label{eq::CI}
\check{\beta}_j \pm z_{\alpha/2}\eta_j \left\{\check{\Sigma}_{adj}(j,j)\right\}^{1/2},\quad j\in\{0, 1,\ldots,p\},
\end{equation}
where $\eta_j = \min\{n^{1/2}\lambda,\max\{1,\lambda/\lvert\hat{\beta}_j\rvert\}\}$ is the adjustment weight, and $z_{\alpha/2}$ is the upper $\alpha/2$ quantile of the standard normal distribution. 
Theorem \ref{prop::moments} below reveals the property of the proposed interval (\ref{eq::CI}).

\begin{theorem}
\label{prop::moments}
Consider the posterior distribution under either the AL prior (\ref{eq::priorAlasso}) or the CA prior (\ref{eq::priorSCAD}). Under the conditions of Theorem \ref{thm::Unified}, we have the following results.
\begin{enumerate}
  \item Convergence of the posterior mean:
  \begin{eqnarray*}
     n^{1/2}(\check{\bbeta}_{\setS} -\bbeta_{\setS}^0) &\rightarrow&  \mathrm{N}\left\{0\,,\,\tau(1-\tau)G_{\setS}^{-1}D_{\setS}G_{\setS}^{-1}\right\}, \\
     n\lambda(\check{\bbeta}_{\setS^c} - \bm{0}) 
&\rightarrow& 0,
\end{eqnarray*}
in distribution as $n \to \infty$.
\item Properties of the adjusted variance:
\begin{eqnarray*}
&n\,\check{\Sigma}_{adj,\setS} & \; = \;  \tau(1-\tau)G_{\setS}^{-1}D_{\setS}G_{\setS}^{-1} + \op(1),\\
(n^{1/2}\lambda)^{-2}\;\lesssim_{\Pstar}\;&(n\lambda)^2\,\check{\Sigma}_{adj}(j,j)&\; \ll_{\Pstar} \; 1,\quad j\not\in \setS.
\end{eqnarray*}
\end{enumerate}
\end{theorem}

Theorem \ref{prop::moments} informs us of several aspects of the posterior inference. First, the posterior mean for the active coefficient is first-order equivalent to the oracle quantile regression estimator as if we knew the set $\setS$. Furthermore, the adjusted posterior variance-covariance matrix captures the sampling variance-covariance of the posterior mean. For those coefficients, the adjustment weight $\eta_j =1 + \op(1)$ due to $n^{-1/2} \ll \lambda\ll 1$, so the confidence intervals in the form of (\ref{eq::CI}) can be viewed as standard Wald-type intervals in the oracle model.

Next we consider any inactive coefficient $\beta_j$ for $j\not\in\setS $. In this case, $\hat{\beta}_j = \Op(n^{-1/2})$, so the adjustment weight $\eta_j \asymp_{\Pstar} n^{1/2}\lambda \to \infty$ in (\ref{eq::CI}) works to inflate the Wald-type interval. Theorem \ref{prop::moments} implies
\begin{eqnarray*}
\frac{\check{\beta}_j - 0}{\eta_j\left\{\check{\Sigma}_{adj}(j,j)\right\}^{1/2}} \to 0,\quad
n^{1/2}\;\eta_j\left\{\check{\Sigma}_{adj}(j,j)\right\}^{1/2}\to 0,\quad j \not\in \setS,
\end{eqnarray*}
in $\Pstar$-probability, and therefore, the confidence interval in (\ref{eq::CI}) will achieve a conservative 100\% asymptotic coverage probability but the interval length remains super-efficient at the order of $\op(n^{-1/2})$.

In summary, the proposed intervals (\ref{eq::CI}) are automatically adaptive to the possible sparsity in the model without relying on a dichotomous variable selection step. In a sparse model  ($s<p$), such interval estimates are more efficient than the classical quantile regression inference using all the coefficients. Empirically we will see later that the proposed intervals are less sensitive to tuning than direct quantile regression inference following model selection.

\begin{remark}
We remark on the value of the Bayesian computational framework: Theorem \ref{thm::Unified} implies the posterior variance-covariance matrix approximates $G_{\setS}^{-1}$, which is an essential quantity for inference in quantile regression and otherwise requires non-parametric estimation \citep{yang2016posterior}. We refer the readers to \citet{chernozhukov2003mcmc} for an in-depth discussion of frequentist inference via MCMC.
\end{remark}

\subsection{Optimally-weighted posterior inference}
\label{subsec::weighted}

In the presence of heteroscedasticity, it is well-known from \citet{newey1990semiparametric} that the following optimally-weighted quantile regression estimator is semi-parametric efficient for estimating $\bbeta^0$ when no sparsity is at play:
\begin{equation*}
\hat{\bbeta}^{(w)} = \argmin_{\bm{u}\in\mathbb{R}^{(p+1)}}\;\sum_{i=1}^n \zeta_i\rho_\tau(y_i - \bx_i^{\T}\bm{u}),
\end{equation*}
where $\zeta_i = f_{\epsilon\mid X = \bx_i}(0)$. In a possibly sparse quantile regression model, a natural question is whether we can achieve the optimal semi-parametric efficiency by using only the data on $(X_\setS , y)$, i.e., after `oracle' model selection is attained. The answer is, somewhat surprisingly, negative, because the "optimal" weights $f_{\epsilon\mid X_\setS}(0)$ under the "oracle model" does not capture the full heteroscedasticity in data. Instead, we show that the statistical efficiency can be further improved by the posterior inference using the the optimally-weighted asymmetric Laplace working likelihood:
\begin{equation}
\label{eq::ALlikelihood-weight}
\mathcal{L}^{(w)}(\mathbb{D}_n \mathrel{\mid} \bbeta) \propto 
\exp\left\{ -\sum_{i=1}^n\zeta_i\rho_\tau(y_i - \bx_i^{\T}\bbeta) \right\}.
\end{equation}
Coupling (\ref{eq::ALlikelihood-weight}) with the shrinkage priors in Section \ref{sec::bayes-model}, we obtain the posterior density  $p^{(w)}(\bbeta \mathrel{\mid} \mathbb{D}_n)$, and we denote the posterior mean by $\check{\bbeta}^{(w)}$. The following result gives the sampling distribution of the posterior mean for the active coefficients. 
\begin{proposition}
\label{prop::SCAD-weighted}
Consider the weighted working likelihood (\ref{eq::ALlikelihood-weight}) and either of the prior (\ref{eq::priorAlasso}) or (\ref{eq::priorSCAD}). Under the same conditions in Theorem \ref{thm::Unified}, the posterior mean satisfies:
  \begin{eqnarray*}
     n^{1/2}(\check{\bbeta}^{(w)}_{\setS} -\bbeta_{\setS}^0) &\;\rightarrow\;&  \mathrm{N}\left\{0\,,\,\tau(1-\tau)Q_{\setS}^{-1}\right\},
\end{eqnarray*}
in distribution,
where $Q_{\setS} = \Estar\{X_{\setS}X_{\setS}^{\T}f_{\epsilon\mid X}^2(0)\}$. 
\end{proposition}

On the other hand, if classical quantile regression is applied to $(X_\setS , Y)$ with the inactive covariates left out, the `optimally' weighted quantile regression has an asymptotic variance of $\tau(1-\tau){V}_{\setS}^{-1}$ \citep{newey1990efficient}, where $V_\setS = \Estar\{X_\setS X^{\T}_\setS f^2_{\epsilon\mid X_{\setS}}(0)\}$ relies only on the active covariates. We show in  the Supplementary Materials that
\begin{equation}
Q_{\setS}^{-1} \preceq  V_{\setS}^{-1},\label{eq::efficiency-compare}
\end{equation}
which reveals that focusing only on the oracle quantile regression model (even when it is available) does not lead to optimal efficiency for the active coefficients. 

There is a simple reason why the inactive set of covariates should not be abandoned. Even though $X_{\setS^c}$ does not affect the conditional $\tau$th quantile of $Y$ given $X$, it may still impact other aspects of the conditional distribution of $Y$ given $X$, in particular the density function $f_{\epsilon\mid X}(0)$ may depend on $X_{\setS^c}$. Unless  $f_{\epsilon\mid X}(0) = f_{\epsilon\mid X_{\setS}}(0)$, the optimal efficiency of quantile regression analysis cannot be achieved if we only focus on those active covariates. In general, a truly `oracle' model should also identify covariates that affect the conditional density function $f_{\epsilon\mid X}(0)$, in addition to $X_{\setS}$.


\begin{remark}
To focus on the main idea, we suppose that the optimal weight $\zeta_i = f_{\epsilon \mid X = \bx_i}(0)$ in (\ref{eq::ALlikelihood-weight}) is known. In practice, it is possible to use the estimated weights while still achieve the same asymptotic efficiency as if we knew $\zeta_i$; see, e.g., \citet{newey1990efficient,koenker1994estimatton} and \citet{zhao2001asymptotically} for some theoretical investigations.
 \end{remark}

\section{Posterior asymptotics with increasing dimensions}
\label{sec::increase-p}

In this Section, we present some 
extensions of the results in Section \ref{subsec::theory}
when the dimension $p=p_n$ diverges with, while still of smaller order than, the sample size $n$; We also allow the size of the active covariates, $|\setS| = s_n$, to depend on the sample size. For illustration purposes, we only focus on the CA prior (\ref{eq::priorSCAD}) in this Section, and we show that the posterior distribution still achieves adaptation to sparsity, even in the regime of moderately increasing dimensions.

The asymptotic regime with an increasing dimension is often of practical interest. 
When modeling the conditional quantile function, it is common to consider Model (\ref{eq::truemodel}) where the complexity may depend on the available sample size. A leading example is when we approximate the unknown conditional quantile function by a linear combination of series/basis expansions, e.g., B-splines, polynomials, and wavelets \citep{chao2017quantile,belloni2019conditional}. To control the approximation error, the number of basis function typically increases with the sample size at a certain rate \citep{he1994convergence}.
The regime also covers the so-called `many regressors' model in econometrics, where a large number of variables are often necessary to model economic theories \citep{cattaneo2018inference}. 

We first discuss some generalizations of the technical assumptions in Section \ref{subsec::theory} when the dimension $p_n \to \infty$. With $p_n = o(n)$, Assumptions \ref{aspt::id} and \ref{aspt::density} are standard in the quantile regression literature \citep{belloni2019conditional,pan2020multiplier}.
On the other hand, Assumptions \ref{aspt::covar} and \ref{aspt::sparsity} may not be suitable for the increasing dimensional regime, therefore we make the the following substitutions for them. 

\begin{asptref}{aspt::covar}[Covariates]
There exists a constant $\sigma_0 > 0$, such that for all $\lVert u\rVert = 1$ and $t>0$:
\begin{equation}
\Pstar\left(|u^TD^{-1/2}X| \geq \sigma_0 t\right)\leq 2\mathrm{e}^{-t}.
\label{eq::sub-exponential}
\end{equation}
Furthermore, the eigenvalues of the matrix $D = \Estar[XX^T]$ satisfies
\begin{equation}
p_n^{-1}\lesssim\theta_{\min}(D) \leq\theta_{\max}(D)\lesssim p_n,\quad\text{and}\quad \theta_{\min}(D_{\setS}) \geq \theta_{1} > 0,
\label{eq::modify-eigen}
\end{equation}
for some constant $\theta_1 > 0$.
\label{aspt::modify-covar}
\end{asptref}


\begin{asptref}{aspt::sparsity}[Sparsity]
There exists a sequence $\underline{b}_n > 0$ such that 
\[
\min_{j\in\setS\setminus \{0\}}\lvert \beta_{j}^0\rvert  > \underline{b}_n.
\]
\label{aspt::modify-sparsity}
\end{asptref}

Assumption \ref{aspt::modify-covar} consists of two parts:
First, (\ref{eq::sub-exponential}) 
states that the standardized covariate $D^{-1/2}X$ is sub-exponential, which strengthens the boundedness of $X$ in Assumption \ref{aspt::covar}; We refer to \citet[Section 3.3]{vershynin2018high} for examples of sub-exponential distributions in high-dimensions. Second, (\ref{eq::modify-eigen}) relaxes Assumption \ref{aspt::covar} by allowing some eigenvalues to vanish or diverge as $p_n\to\infty$, implying that there could be some degree of co-linearity among the $p = p_n$ covariates. 
Finally, Assumption  \ref{aspt::modify-sparsity} requires all non-zero coefficients to be sufficiently separated from $0$, yet the threshold $\underline{b}_n$ is allowed to shrink towards zero as the sample size grows.

The result below generalizes Theorem \ref{thm::Unified} to an increasing dimensional regime, where we drop the subscript $n$ in $s$ and $p$ for simplicity. 

\begin{theorem}
Consider the posterior distribution under the CA prior (\ref{eq::priorSCAD}) and $p\to\infty$. Suppose Assumptions \ref{aspt::id}, \ref{aspt::modify-covar}, \ref{aspt::density} and \ref{aspt::modify-sparsity} hold. If  $s^4p^2\log^2 n = o(n)$, and the tuning parameter $\lambda$ is chosen such that 
\begin{equation}
\frac{s^{1/2}p\log^{3/2} p}{n^{1/2}} \ll \lambda \ll \min\left\{s^{-1/2},\;\underline{b}_n,\;\underline{b}_n[\theta_{\min}(D)]^{1/2}\right\},
\label{eq::lambda-in-p}
\end{equation}
then we have the following results:
\begin{enumerate}
\item Adaptive rate-of-contraction: for any sequence $M_n\to+\infty$.
\[
\Pi\left( \left\lVert\bbeta_{\setS} - \bbeta^0_{\setS}\right\rVert \leq M_n\sqrt{\frac{s}{n}} \,,\, \left\lVert\bbeta_{\setS^c} \right\rVert_{\infty} \leq M_n\frac{s\log p}{n\lambda} ~\middle\vert~ \mathbb{D}_n\right) \to 1,
\]
in $\Pstar$-probability
\item Distributional approximation: for $\pi_j(u) = (1/2)\exp\{-|u|\}$, $(\forall j\in\setS^c)$,
\[
\left\lVert p\left(\bbeta ~\middle\vert~  \mathbb{D}_n\right) -  \phi\left(\bbeta_S;\tilde{\bbeta}_S,\frac{1}{n}G_{\setS}^{-1}\right)\times \prod_{j\not\in\setS} \left\{ n\lambda  \pi_j(n\lambda \beta_j)\right\}\right\rVert_{TV} \to 0,
\]
in $\Pstar$-probability, where $\phi(\cdot\,;\,\mu,\Sigma)$ is the density function of a multivariate-Gaussian distribution; $\tilde{\bbeta}_{\setS}$ and $G$ are defined in Section \ref{sec::notations}.
\end{enumerate}
\label{thm::SCAD-in-p}
\end{theorem}

Theorem \ref{thm::SCAD-in-p} explicitly characterizes the effect of increasing model dimension on the posterior. Since $(n\lambda)/(s\log p) \gg (np)^{1/2}$, part 1 of Theorem \ref{thm::SCAD-in-p} shows the posterior distribution for all inactive coefficients concentrates simultaneously towards zero at a second-order rate, despite that there may be a diverging number of them. 
For part 2 of Theorem \ref{thm::SCAD-in-p}, it is sometimes more informative to consider a one-dimensional linear combination of parameters $\bm{\alpha}^T\bbeta$ for $\lVert \bm{\alpha} \rVert = 1$ in the regime of increasing dimension \citep{fan2004nonconcave}.
 If $\bm{\alpha}_{\setS}\neq \bm{0}$, then the posterior for $\bm{\alpha}^T\bbeta$ would be asymptotically `oracle'; otherwise the posterior would have a scale at the order of $p^{1/2}/(n\lambda) \ll n^{-1/2}$, which is super-efficient. 

While Theorem \ref{thm::SCAD-in-p} covers a wide range of models under general design and spasity conditions, the range (\ref{eq::lambda-in-p}) may imply additional conditions on $\underline{b}_n$, $p_n$, or the eigenvalues of $D$ in a given setting. To better explain the conditions in Theorem \ref{thm::SCAD-in-p}, here we consider an example with a sparse model, i.e., $s_n = s_0$ stays fixed yet $p_n\to\infty$. In addition to Assumptions \ref{aspt::id}, \ref{aspt::modify-covar}, \ref{aspt::density} and \ref{aspt::modify-sparsity}, we suppose the design matrix satisfy $\theta_{\min}(D) \geq \theta_0 > 0$, which aligns with the setting in \citet{belloni2019conditional}.
Under this model setting, the conclusions in Theorem \ref{thm::SCAD-in-p} hold if
\begin{align*}
&p^2\log^2 n = o(n),\quad\underline{b}_n\gg \frac{p\log^{3/2} p}{n^{1/2}},\\
&\text{and}\qquad
\frac{p\log^{3/2} p}{n^{1/2}} \ll \lambda_n \ll \underline{b}_n,
\end{align*}
where $\underline{b}_n$ is defined in Assumption \ref{aspt::modify-sparsity}.
With a sparse model, the above conditions are more intuitive and are comparable with the literature on shrinkage estimation with moderately increasing dimensions \citep{fan2004nonconcave,huang2008adaptive,armagan2013generalized}, even though we work with a mis-specified likelihood.

\begin{remark}
For high-dimensional sparse problems with $s_n \ll n \ll p_n$, it is often more practical to employ a screening step first to reduce the dimension to a manageable scale \citep{fan2008sure}, and then pursue statistical inference via the Bayesian approach. For example, the Sure Independence Screening for quantile regression in \cite{he2013quantile} or \citet{wu2015conditional} can select a $d_n$-dimensional sub-model, such that $d_n = o(n)$ but all active covariates are retained with probability approaching one. 
Our asymptotic regime in this Section then becomes relevant if we focus on the selected model, and Theorem \ref{thm::SCAD-in-p} applies to the $d_n$-dimensional posterior distribution post-screening.
\end{remark}


\section{Simulation}
\label{sec::simulation}

We use a Monte Carlo simulation to demonstrate that the asymptotic properties established in this paper are visible in finite-sample problems. A limited comparison with some other inferential methods in quantile regression is also included. We only highlight several key findings here, whereas more detailed results are relegated to the online Supplementary Materials.

We generate random samples of size $n$ from the following regression model
\[
Y = 1 + 3X_2 - 5X_5 +\left\{\frac{1+ (X_6 - 1)^2}{3}\right\}  e,
\]
where $e\sim \mathrm{N}(0,1)$ is independent of the covariate vector $X = (X_1,\ldots,X_6)^T \sim\mathrm{N}(0,\Sigma)$ with the $(i,j)$th entry of $\Sigma$ being $0.8^{|i-j|}$ for $i,j \in \{1, \cdots , 6\}$. The data generating process satisfies Model (\ref{eq::truemodel}) at $\tau = 0.5$, where $X_2$ and $X_5$ are active but $X_6$ is inactive for the conditional median of $Y$ given $X$.  We consider two different sample sizes $n = 200$ and $n = 500$, and use $2,000$ Monte Carlo data sets in each simulation.

We compare the proposed posterior inference method with four other approaches for constructing $90\%$ confidence intervals of the median regression coefficients. Three of competing approaches are the robust rank-score method of \citet{koenker1999goodness} applied to: (i) the full model with $(X_1,\ldots,X_6)$ included, (ii) the oracle model with $(X_2,X_5)$ included, and (iii) the selected model from adaptive lasso variable selection, respectively. 
The fourth competing approach is the wild bootstrap for the adaptive lasso quantile regression proposed recently by \citet{wang2018wild}. Not all possible methods are included in this study, but those competitors are known to have generally good performance under heteroscedastic models.

We first compare the performances of those methods under a fixed tuning parameter in Table \ref{tab::qid}. To make a fair comparison, the tuning parameter $\lambda$ for both the shrinkage prior and the adaptive lasso model selection are kept the same across all Monte Carlo data sets at a given sample size. We relegate further implementation details, including tuning parameter specification, to the online Supplementary Materials.

\renewcommand{\arraystretch}{0.5}
\begin{table}[t!]
\begin{threeparttable}
\caption{Empirical coverage probabilities and average lengths ($\times 100$) for $90\%$ confidence intervals.}%
\begin{tabular}{lcccccc}
\toprule
 &\multicolumn{3}{c}{Empirical coverage}& \multicolumn{3}{c}{Average length (s.e.)}\\
 \midrule
 \multicolumn{7}{c}{$n = 200$}\\
 &$\beta_2$&$\beta_5$&$\beta_{zeros}$&$\beta_2$&$\beta_5$&$\beta_{zeros}$\\
 \midrule
Full &  92 &  91 & 90 & \hphantom{00}43.7 (0.25) & 43.9 (0.25) & 41.7 (0.10) \\ 
  Oracle &  89 &  93 & 100 & \hphantom{00}22.5 (0.13) & 32.6 (0.21) &  0.0 (0.00) \\ 
  Refit &  84 &  86 &  89 & \hphantom{00}28.1 (0.21) & 34.7 (0.22) & 11.4 (0.06) \\ 
  WildPen &  85 &  84 &  89 &\hphantom{00} 27.0 (0.13) & 30.2 (0.14) & 20.1 (0.08) \\ 
  BayesAdj &  93 &  93 &  96 & \hphantom{00}28.1 (0.11) & 32.7 (0.12) & 11.7 (0.04) \\ 
  \midrule
  \multicolumn{7}{c}{$n = 500$ }\\
  &$\beta_2$&$\beta_5$&$\beta_{zeros}$&$\beta_2$&$\beta_5$&$\beta_{zeros}$\\
\midrule
    Full &  91 &  91 &  90 &\hphantom{00} 26.7 (0.12) & 26.6 (0.12) & 25.6 (0.06) \\ 
  Oracle &  90 &  93 & 100 & \hphantom{00}14.0 (0.06) & 20.5 (0.11) &  0.0 (0.00) \\ 
  Refit &  81 &  86 &  89 & \hphantom{00}17.2 (0.11) & 21.5 (0.11) &  6.3 (0.05) \\ 
  WildPen &  84 &  85 &  91 &\hphantom{00} 16.7 (0.07) & 18.9 (0.07) & 11.3 (0.06) \\ 
  BayesAdj &  89 &  91 &  95 &\hphantom{00} 16.3 (0.06) & 19.3 (0.06) &  6.1 (0.03) \\
  \bottomrule
\end{tabular}
\label{tab::qid}
\begin{tablenotes}
\linespread{1}\footnotesize
\item 'Full' refers to the rank-score method applied to all the covariates, 'Oracle' uses only the active covariates for the conditional median, and 'Refit' is the rank-score method applied to a model selected by adaptive lasso.
'WildPen' is the wild bootstrap approach of \citet{wang2018wild}.
'BayesAdj' refers to the proposed adjusted posterior inference in Section \ref{subsec::inference}.
For the 'Refit' and 'Oracle' methods, if a covariate is not included in the model, we report its confidence interval as a singleton $\{0\}$.
The column $\beta_{zeros}$ averages over all inactive coefficients $\beta_1$, $\beta_3$, $\beta_4$ and $\beta_6$.
The numbers shown in the parentheses are the estimated standard errors. For the coverage estimates, their standard errors are all below $0.9$. For penalization/shrinkage, we used $\lambda = 0.066$ when $n=200$ and $\lambda = 0.051$ when $n = 500$.
\end{tablenotes}
\end{threeparttable}
\end{table}

Table~\ref{tab::qid} suggests that the adjusted posterior inference indeed achieves adaptive performance.
For the active coefficients, the adjusted posterior inference gives much shorter intervals than those from  the full model, and it is reasonably competitive with the results from the oracle model.
For the inactive coefficients, the adjusted posterior inference gives much shorter intervals than those under the full model with higher-than-nominal coverage probability. 
On the other hand, the  wild bootstrap approach and the rank-score method applied to the selected model from adaptive lasso both fall short in coverage. Part of the reason for their under-performance is that the adaptive lasso does not achieve 'oracle' selection often enough in this case due to limited sample size, even when $n = 500$; See Figure S2 in the Supplementary Materials. Therefore, those approaches based on variable selection may fail to fully account for the uncertainty induced by selection.

In addition, the adjusted posterior inference gives more stable confidence intervals, as the standard errors for average interval lengths are among the smallest of all methods in Table \ref{tab::qid}. Such finite-sample stability of the adjusted posterior inference reflects its avoidance of pursuing dichotomous variable selection. We refer to Figure S1 and comments thereof in the Supplementary Materials for more details.

Next, we examine the impact of the tuning parameter in the comparisons of shrinkage-based methods. To this end, we vary $\lambda$ at a wide range of values and compare the performances in Figure \ref{fig_sensitivity} when sample size $n = 500$; see also Figure S3 in the Supplementary Materials for the results when $n = 200$. 
We note that the coverage probabilities of the adjusted posterior inference method for the active coefficients are more stable around the nominal levels than other methods for a wide range of $\lambda$ values.
For the inactive coefficient $\beta_6$, the coverage probability for the proposed method remains high without any sacrifice in the lengths of the intervals relative to other non-oracle methods. Finally, we note from our empirical studies that the adjusted posterior inference tends to lose coverage if the shrinkage parameter $\lambda$ is too large. As a practical guide, we suggest choosing a $\lambda$ that is slightly smaller than what one would obtain from the cross-validation method for the adaptive lasso.

\begin{figure}[t!]
\centering
         \includegraphics[width = 0.85\textwidth]{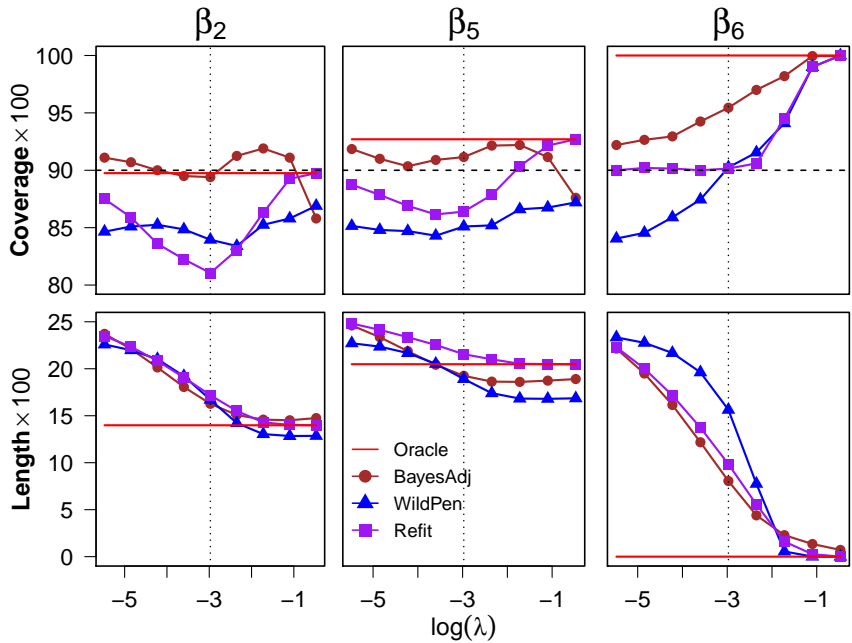}\vspace{-0.5em}
\caption{Empirical coverage probabilities and average lengths for 90\% confidence intervals with different $\lambda$ when $n = 500$. The true regression coefficients are $\beta_2^0 = 3$, $\beta_5^0 = -5$ and $\beta_6^0 = 0$. The value of $\lambda$ marked by a vertical broken line is used to produce Table ~\ref{tab::qid}, and the abbreviated method names are the same as Table~\ref{tab::qid}. }
\label{fig_sensitivity}
\end{figure}

\section{Conclusion and discussion}
\label{sec::conclude}

In this paper, we show that Bayesian computational framework can be useful for constructing frequentist confidence intervals in possibly sparse quantile regression analysis.
By employing appropriate shrinkage priors, we show the posterior inference can adapt automatically to model sparsity. Asymptotically, the proposed confidence intervals are oracle efficient for the active coefficients, and are super-efficient for the inactive coefficients.
The posterior inference approach enjoys two distinct advantages: 
(i) it avoids the need to pursue dichotomous variable selection by employing a continuous shrinkage prior;
(ii) it offers an alternative to classical frequentist computation by trading nuisance-parameter estimation for MCMC sampling. The proposed approach demonstrates desirable finite-sample stability in simulation studies.

We highlight a few possible extensions of our work. 
In the paper, we demonstrate that the asymptotic adaptation phenomena of the posterior in problems of moderately increasing dimensions.
Nonetheless, it remains an interesting problem to understand the value of posterior inference in the high-dimensional setting where $p \gg n$.
In addition, exploiting the Bayesian computational methods would be even more appealing in more complex settings, e.g., censored quantile regression \citep{yang2016posterior,wu2021bayesian} where the objective function can be highly non-convex \citep{powell1984least,powell1986censored}. The Bayesian approach can avoid direct optimization of the objective function while incorporating possible model sparsity.

\section*{Supplementary Materials}

The online supplementary material contains some additional results from the simulation study, as well as the proofs of all the results in this paper.
\par
\section*{Acknowledgements}
The work is partially supported by the NSF Awards DMS-1914496 and DMS-1951980.
\par


\bibhang=1.7pc
\bibsep=2pt
\fontsize{9}{14pt plus.8pt minus .6pt}\selectfont
\renewcommand\bibname{\large \bf References}
\expandafter\ifx\csname
natexlab\endcsname\relax\def\natexlab#1{#1}\fi
\expandafter\ifx\csname url\endcsname\relax
  \def\url#1{\texttt{#1}}\fi
\expandafter\ifx\csname urlprefix\endcsname\relax\def\urlprefix{URL}\fi

 \bibliographystyle{chicago}      
 \bibliography{QR_SS}   

\vskip .65cm
\noindent


\end{document}